\documentclass[journal,comsoc]{IEEEtran}

\usepackage{cite}

\ifCLASSINFOpdf
\else
\usepackage[dvips]{graphicx}
\fi
\usepackage{amsmath}
\usepackage{newtxmath}

\usepackage{array}

\usepackage{fixltx2e}
\usepackage{multirow}

\begin{document}

\title{Low-Complexity Null Subcarrier-Assisted OFDM PAPR Reduction with Improved BER}

\author{Md~Sakir~Hossain,~\IEEEmembership{ Graduate Student Member,~IEEE,}
        and~Tetsuya~Shimamura,~\IEEEmembership{Member,~IEEE}
\thanks{Manuscript received xx, 2016; revised xx, xx.}
\thanks{M. S. Hossain was with the Graduate School of Science and Engineering,  Saitama University, Saitama City 338-8570, Japan
 (e-mail: sakir@sie.ics.saitama-u.ac.jp).}
\thanks{T. Shimamura is with the Information Technology Center, Saitama University, Saitama City 338-8570, Japan (e-mail:shima@sie.ics.saitama-u.ac.jp)}
}

\markboth{IEEE COMMUNICATION LETTERS,~Vol.~xx, No.~xx, xx~xx}%
{Shell \MakeLowercase{\textit{et al.}}: Bare Demo of IEEEtran.cls for IEEE Journals}

\maketitle

\begin{abstract}
In this letter, we propose a low-complexity data-null subcarrier switching-based peak-to-average power ratio (PAPR) reduction scheme for the orthogonal frequency division multiplexing (OFDM) systems, which provides improved bit error rate. We perform the switching between the data and null subcarriers in such a way that distance between any two switched null subcarriers remains constant. This method can achieve up to 6 dB signal-to-noise ratio gain and shed the system complexity by more than 98\% of the conventional scheme, with a slight compromise of the PAPR reduction capability.
\end{abstract}


\IEEEpeerreviewmaketitle

\section{Introduction}
\IEEEPARstart{O}{rthogonal} frequency division multiplexing (OFDM) has been seen as a robust signal processing technique due to its high spectral-efficiency and immunity to intersymbol interference in multi-path fading.  Inspired by these advantages, a number of wireless system standards, such as IEEE 802.11, IEEE 802.16, IEEE 802.22 and so on, have adopted OFDM as their physical layer (PHY) technology. Its inclusion in the forthcoming fifth generation (5G) wireless system, however, faces great challenges from the other candidate PHY technologies due to its low energy efficiency caused by high peak-to-average power ratio (PAPR). High PAPR occurs when several sinusoids get added coherently in inverse fast Fourier transform (IFFT) at a particular instant. The resulting peaks which exceed the operating region of a power amplifier (PA) cause nonlinear distortion at the output of the PA. This nonlinearity causes bit error rate (BER) degradation, in-band and out-of-band (OOB) interferences.

There are a plethora of solutions~\cite{rahmat}, which have been proposed to solve the longstanding problem. Most of the solutions are spectrally-inefficient because a certain portion of available bandwidth is allocated for this purpose, thereby reducing system throughput. Very few works have been done to date to devise a PAPR reduction scheme without compromising spectral-efficiency. Among them, the recently proposed null subcarrier-assisted PAPR reduction scheme~\cite{wong1} has drawn a considerable attention of the researchers. In such schemes, some of the null subcarriers are used as a tool to reduce PAPR; the positions of certain number of null-subcarriers are switched with all data subcarriers iteratively and the combination that provides the lowest PAPR is transmitted. Instead of sending any side information (SI) to let the receiver know the locations of the switched null-subcarriers (SNSs), a blind SNS detection is carried out at the receiver based on the zero-energy property of the null subcarriers. This technique reduces PAPR significantly, requires no SI, and causes no loss of orthogonality like the well-known clipping technique. However, it suffers from its extremely high computational complexity and poor BER at low signal-to-noise ratio (SNR) due to inaccurate SNS detection. To reduce its prohibitively high computational overhead, few works are done~\cite{sabbir1} -\cite{sabbir3}. All these methods can shed a significant part of the overall complexity. The resulting remaining complexity is, however, still very high compared to the other existing PAPR reduction techniques. To reduce the complexity, the authors of the original scheme~\cite{wong1} propose a technique in~\cite{wong2}, where the main attractive feature of the null subcarrier-assited technique~\cite{wong2}, no SI requirement, is sacrificed. Such an attempt makes this technique unattractive because the other existing SI-required techniques can reduce more PAPR with much lower computational complexity. In addition to the high computational complexity, the poor SNS detection of the blind method escalates BER. To the best of our knowledge, no work has yet been done to address this problem. 

In this letter, our goal is to find an efficient switching technique, which will attain a significant PAPR reduction in one hand, and reduce the system complexity and improve BER on the other hand.  We propose to switch the positions of the data and null subcarriers in such a way that distance between any two consecutive SNSs can be kept constant. This sheds more than 98\% computational burden of the original method~\cite{wong1}. At the receiver, we use this property in addition to the conventional zero-energy property of the null subcarriers for efficient detection of SNSs. This increases detection accuracy rate, thereby improving the BER performance.

\section{Conventional Null subcarrier based PAPR Reduction}

In wireless system standards like WiMax (IEEE 802.16) and wireless LAN (IEEE 802.11), a small percentage of the available subcarriers are placed at the center and both sides of data subcarriers. Such subcarriers are called null subcarriers. They carry no information and are employed to form spectrum and reduce OOB radiation. In the implementation of OFDM, they are zero-energies (that is, zero is assigned to the null subcarriers instead of unity energy of phase shift keying (PSK) for example) in the frequency domain. At the receiver, the frequency domain null subcarriers are separated depending on their known locations. 

According to \cite{Li}, it is tolerable to use some of the null subcarriers for other purposes. Inspired by this, a null subcarrier-assisted PAPR reduction scheme is proposed in~\cite{wong1}. To show how this technique works, let us consider two sets: $D$ of length $N_d$ and $G$ of length $N_g$, which consist of indexes of the data and null subcarriers, respectively, and $N=N_d+N_g$; the indexes are in ascending order. If the inner-most $P$, which is an even integer, out of $N_g$ null-subcarriers are selected for switching with $P$ unknown data subcarriers, the set $G_s=\{g_{s_1},g_{s_2},...,g_{s_P}\}$   contains the indexes of the $P$ pre-selected null subcarriers. The $P$ out of $N_d$ data subcarriers are selected randomly for interchanging positions with the $P$ pre-selected null subcarriers. This selection can be done in $N_d \choose P$ $=\frac{N_d!}{P!(N_d-P)!}$ different ways. For each $P$ data subcarriers selection, the selected null and data subcarriers are switched following the switching constraint described below and PAPR of the corresponding time domain signal is computed; thus, there will be $N_d \choose P$ time domain signals and their corresponding PAPRs. The lowest PAPR time domain signal is transmitted. Suppose that the set $D_s=\{d_{s_1},d_{s_2},...,d_{s_P}\}$    contains the indexes (in ascending order) of the $P$ switched data subcarriers of the transmitted signal. The switching between the data and null subcarriers is done in such a way that if    $d_{s_m}<d_{s_m+1}$, then $g_{s_m}<g_{s_m+1}$, we will call this constraint as switching constraint in the rest of this paper. Using this switching constraint and zero-energy property of the null subcarriers, the de-switching is performed at the receiver without any kind of SI. 

\section{Proposed Scheme}
\begin{figure}[t]
\centering
\includegraphics[width=3.35in]{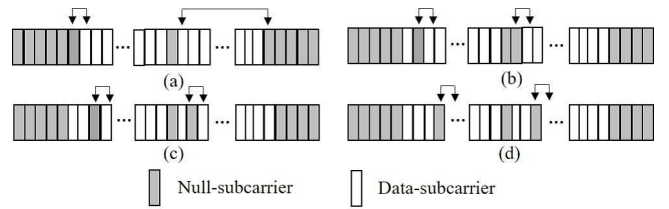}
\caption{Switching between the data and null subcarriers.}
\label{fig:fig1}
\end{figure}

In this letter, we propose a null subcarrier based new blind PAPR reduction scheme to improve BER and computational complexity of the scheme proposed in \cite{wong1}. Instead of swapping the positions of the randomly-selected data subcarriers with the pre-selected set of null subcarriers, we plan to interchange the positions of these two kinds of subcarriers in a predetermined manner so that an equidistance property can be maintained, where the equidistance property is a characteristic of having a constant distance between any two consecutive SNSs. This constant distance is defined by $R=\frac{N_d}{P}$. Usually, $\frac{N_g}{2}$  and $\frac{N_g}{2}-1$ out of $N_g$ null subcarriers are placed to the left and right of the data sequence, respectively, and the rest one is placed at the center of it. Fig.~\ref{fig:fig1} shows the subcarrier arrangement and switching procedure between the data and null subcarriers, where the double arrows indicate the data and null subcarriers between which switching will be done in this phase of switching. The conventional arrangement of the subcarriers is shown in Fig.~\ref{fig:fig1}(a) for $P=2$. In this case, the indexes of all subcarriers range from 1 to $N$, where 1 to $\frac{N_g}{2}$, $\frac{N_d}{2}+\frac{N_g}{2}+1$, and $N-(\frac{N_g}{2}-1)$ to $N$ are null subcarriers' indexes and the rest are the indexes of the data subcarriers. The inner-most $P$ (with $\frac{P}{2}$ from each side) null subcarriers are selected for switching. The first phase of switching is done using the following rule: for any positive integer  $K\leq \frac{P}{2}$ , the $N-N_d-\frac{N_g}{2}-(\frac{P}{2}-K)$–th subcarsubrier is switched with the $\frac{N_g}{2}+\frac{(K-1)N_d}{P}+1$-th subcarrier, and when $K>\frac{P}{2}$, the switching is done between the $N-(\frac{N_g}{2}-1)+(K-\frac{P}{2})$-th and $\frac{N_g}{2}+\frac{(K-1)N_d}{P}+2$-th subcarriers. In the second phase of switching (see Fig.~~\ref{fig:fig1}(b)), each SNS swaps its position with the data subcarrier located just to right. The resulting combination of the data and null subcarriers is shown in Fig.~\ref{fig:fig1}(c); in this phase, the switching is done following the rule used in the second phase, that is, the SNS with an index $s_m$ interchanges its position with the data subcarrier having the index $s_m+1$. This process of switching continues until the index of the last SNS is less than $N-(\frac{N_g}{2}-1)$ or the allowable maximum iteration is met. PAPR of the time-domain version of each of these combinations is computed, and the combination which provides the lowest PAPR is selected for transmission.

   We will now discuss the SNS detection technique. All conventional blind data-null subcarrier switching based systems~\cite{wong1,sabbir1,sabbir2,sabbir3} use the same SNS detection technique, which can be described as follows. As mentioned before, zero is assigned to null subcarriers in the frequency domain by the transmitter~\cite{Li}. At the receiver, the null detection is carried out in the frequency-domain signal.  In case of a noise free channel, the subcarriers which have zero energy are identified as the SNSs. Since noise free channels are not available in practice, the SNSs possess some noise power. For this reason, the $P$ out of $N_d$ ($N_d-P$ data subcarriers and $P$ SNSs) subcarriers having the lowest power are considered as the SNSs. This searching is carried out among the subcarriers having indexes from $\frac{N_g}{2}+1$ to $\frac{N_g}{2}+\frac{N_d}{2}$ and $\frac{N_g}{2}+\frac{N_d}{2}+2$ to $N-\frac{N_g}{2}$. Upon the detection, the de-switching is done with the SNSs using the switching constraint (i.e., if $d_{s_m}<d_{s_m+1}$, then $g_{s_m}<g_{s_m+1}$). The detection accuracy is, however, very low, which results in extremely poor BER. Since all existing schemes~\cite{wong1,sabbir1,sabbir2,sabbir3} use this detection technique, they suffer from the same BER performance.

   To improve the detection accuracy, we will use the equidistance property in addition to the zero-energy property. The received time-domain signal is first converted to its corresponding frequency-domain version applying fast Fourier transform (FFT). The proposed SNS detection method is applied on this signal. Since all the SNSs are placed in the data subcarriers maintaining a predefined distance from each other, firstly all possible sets of SNSs' indexes are generated. Let us consider a matrix $S$ of dimension $R\times P$, where each element of $S$ is defined in the following way:
 \begin{equation*}\label{eq.1}
  S_{i,j}=
  \begin{cases}
  u_i+nR & \text{for  $0\leq n<\frac{P}{2}$}\\
  u_i+nR+1 & \text{for $P/2\leq n< P$}
  \end{cases}
  \end{equation*}
 where $U=[u_1\;u_2\;u_3\;.\;.\;.\;u_R]^T=[\frac{N_g}{2}+1\;\frac{N_g}{2}+2\;\frac{N_g}{2}+3\;.\;.\;.\;\frac{N_g}{2}+R]^T$, and $n$ is an integer ranging from 0 to $P-1$. Each row of $S$ represents a candidate set for indexes of the SNSs. For each candidate set, the total power of the subcarriers whose indexes are contained in that specific row of $S_i$ is computed. The candidate set which provides the lowest total power among these $R$ sets of candidates is selected as the set of the indexes of the true SNSs. Then the de-switching between the SNSs and switched data subcarriers is done in the usual manner using the switching constraint.
\begin{figure}[!t]
\centering
\includegraphics[trim = 15mm 10mm 5mm 10mm,width=3.75in]{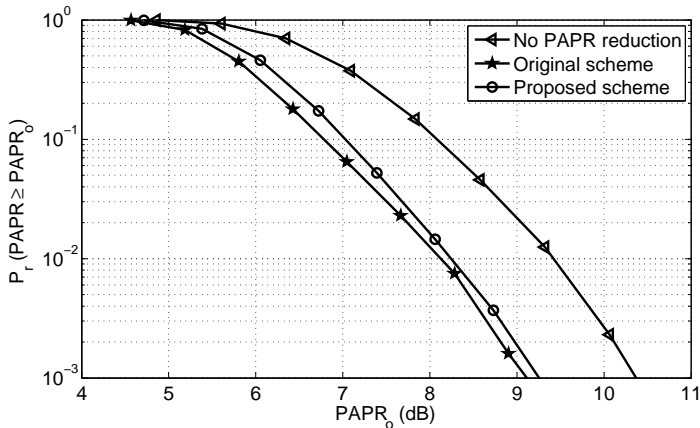}
\caption{PAPR comparison between the conventional and proposed schemes.}
\label{fig2}
\end{figure}

\begin{table}[]
\centering
\caption{SNS detection rate (\%)}
\label{my-label}
\begin{tabular}{|l|l|l|l|l|l|l|}
\hline
\multicolumn{2}{|l|}{SNR (dB)} & 0 & 3 & 6 & 9 & 12 \\ \hline
\multirow{2}{*}{P=2}  & Conventional  & 0.7 & 4.1 & 27.9 & 82.8 & 99.8 \\  \cline{2-7}
                   & Proposed  & 20.5 & 49.2 & 88.3 & 99.8 & 100 \\ \hline
\multirow{2}{*}{P=4}  & Conventional  & 0 & 0.4 & 12.2 & 72.0 & 99.5 \\ \cline{2-7}
                   & Proposed  & 50.9 & 86.3 & 99.6 & 100 & 100 \\ \hline
\end{tabular}
\end{table}

\section{Performance Evaluation}
Performance of the proposed scheme will be analysed in terms of PAPR reduction capability, computational complexity and BER. While the computational complexity will be evaluated analytically, the rest two will be evaluated through simulations. The simulation parameters are as follows: 52 data subcarriers out of 64 total subcarriers like \cite{sabbir3}, 12 null-subcarriers, Quadrature Phase Shift Keying (QPSK) modulation, and four times oversampling. Every simulation result is taken sending $10^4$ OFDM symbols.

   A comparison of complementary cumulative distribution function (CCDF) of PAPR of the proposed scheme to that of the conventional null subcarrier-based PAPR reduction scheme~\cite{wong1} and the original OFDM is shown in Fig.~\ref{fig2}. It is clear that the PAPR reduction capability of the proposed scheme is very close to that of the conventional one; while PAPR of less than 0.1\% of the transmitted symbols of the original OFDM exceed 10.37 dB, the PAPR for the conventional and proposed schemes are 9.11 dB and 9.23 dB, respectively, at the same CCDF level; that is, the proposed scheme degrades 9.5\% of the PAPR reduction capability of the conventional null subcarrier-based PAPR reduction scheme. At CCDF level of $10^{-2}$, this degradation goes down to 7.7\%. In the proposed scheme, the reduction of search space from $N_d \choose P$ to $\frac{N_d}{P}$ decreases the probability of obtaining the lowest PAPR signal; this results in slight rise of PAPR.

   The BER of a null subcarrier-assisted blind PAPR reduction scheme largely depends on the SNS detection accuracy at the receiver. Every wrong detection results in a wrong de-switching between the switched-data and null subcarriers, which ends up BER degradation. The SNS detection comparison between the conventional null subcarrier assisted PAPR reduction scheme and proposed one over additive white Gaussian noise channel in different SNR cases is given in Table 1. As is seen, the proposed scheme far outperforms the conventional one in detecting SNSs. For instance, the accuracies of the conventional scheme are 0.7\% and 0\% for $P=$2 and 4, respectively, at 0 dB SNR. In contrast, the corresponding detection accuracies of the proposed scheme are about 20\% and 51\%. While the detection accuracy of the conventional scheme degrades with the increase of the SNSs, a reverse relation is observed for the proposed scheme; an outstanding improvement is found with larger number of SNSs.
   
      \begin{figure}[!t]
\centering
\includegraphics[trim = 8mm 10mm 5mm 10mm,width=3.75in]{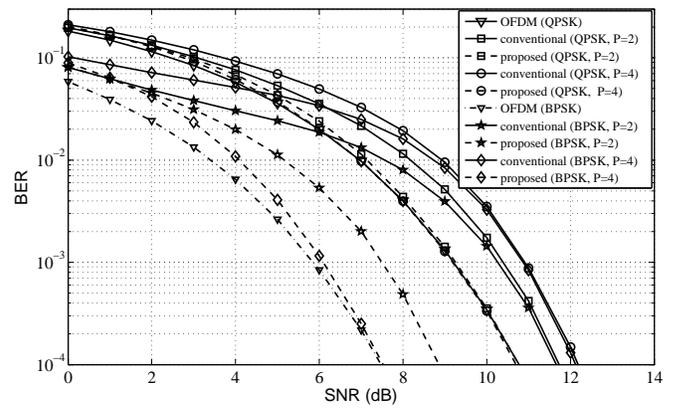}
\caption{BER performance comparison.}

\label{fig3}
\end{figure}
   
\begin{figure}[!t]
\centering
\includegraphics[trim = 12mm 10mm 5mm 10mm,width=3.75in]{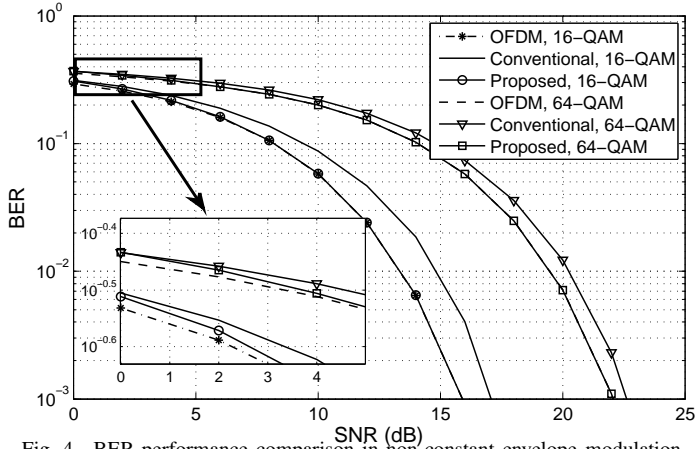}
\caption{BER performance comparison in non-constant envelope modulation}

\label{fig4}
\end{figure}

   The effect of the SNS detection accuracy on the BER is presented in Fig.~\ref{fig3}. To reveal how much BER degradation the blind PAPR reduction scheme itself is responsible for, no PA is included in the transmitter. Fig.~\ref{fig3} reveals that the improvement in SNS detection causes better BER. For $P=4$ and binary phase shift keying (BPSK) modulation, while a target BER of $10^{-3}$ can be achieved by the conventional scheme at 11.85 dB SNR (which is about 6 dB higher than that of the original OFDM), the proposed one can reach the target with a SNR of 6.1 dB (only 0.2 dB higher than that of the original OFDM). For QPSK modulation, the proposed scheme does not cause SNR loss to achieve the target BER for any value of $P$. The conventional scheme, however, requires 1.3 and 1.7 dB more SNR for $P=2$ and $4$, respectively, for the same target BER. Fig.~\ref{fig3} also shows that higher $P$ lets the proposed scheme achieve better BER, although an opposite result is observed for the conventional scheme. Like the performance improvement in PSK, the proposed scheme outperforms the conventional one in non-constant envelope modulation as well. The BER performance for higher-order quadrature amplitude modulation (QAM) is investigated in Fig.~\ref{fig4} for $P=4$. While the conventional scheme requires 1.2 and 0.55 dB more SNR to achieve the target BER of $10^{-3}$ for 16-QAM and 64-QAM, respectively, the proposed scheme does not cause any SNR penalty to attain the same BER for both modulations. It is observed that the SNR gain reduces with the increase of modulation order; this is due to the fact that the effect of closer constellation points of the higher modulation becomes more dominant than the SNS detection rate in determining BER. Since the SNR gain solely depends on the effect of SNS detection rate, this decreasing effect of the detection rate reduces the SNR gain. It is pertinent to mention here that through extensive simulations, the proposed scheme is found to outperform the conventional scheme in the presence of PA as well.

   We will now evaluate the computational complexity of the proposed scheme in comparison to the conventional ones. The number of IFFT operations, PAPR calculations at the transmitter and null subcarrier detection at the receiver contribute a staple part of the total complexity of the system. For this reason, the complexities contributed by these three tasks are considered here and the tasks whose complexities are common for all the methods are not considered. We compare the complexity of our proposed scheme with that of the data-null subcarrier switching based methods which do not require SI (e.g., original method~\cite{wong1} and Ahmed et. al.~\cite{sabbir2}). The number of IFFT operations required by the original method~\cite{wong1} and our proposed one are $N_d \choose P$ and $R$, respectively. To compare the complexity, we will first derive a generalized formula for the maximum number of required IFFT operations, $M_s$, of Ahmed et. al. in~\cite{sabbir2} as follows: 
   \begin{equation*}
   \begin{split}
   M_s&=R+(2R-1)+(3R-2)+(4R-3)+\;.\;.\;.\;\\
   & \quad +[PR-(P-1)]\\
    & =\left(\frac{P^2}{2}+\frac{P}{2}\right)R-\left(\frac{P^2}{2}-\frac{P}{2}\right)\\
    & =\frac{P}{2}[(P+1)R-P+1]
   \end{split}
   \end{equation*}

\begin{table}[]
\centering
\caption{Computational Complexity Comparison}
\label{tabl2}
\begin{tabular}{|c|c|c|}
\hline
Scheme & No. of additions & No. of multiplications \\ \hline
Original & $N_d \choose P$$(Q+4N-1)$ & $N_d \choose P$$\frac{Q+8N+2}{2}$ \\ \hline
Ahmed et. al. & $M_s(Q+4N-1)$ & $\frac{M_s(Q+8N+2)}{2}$ \\ \hline
Proposed & $R(Q+4N-1)+\frac{(4R-2P^{-1}+1)}{2P}$ & $\frac{R(Q+8N+2)}{2}+N_d$ \\ \hline
\end{tabular}
\end{table}

    Each $N$-points IFFT operation requires $\frac{N}{2}log_2(N)$ multiplications and $Nlog_2(N)$ additions. Suppose that $Q=Nlog_2(N)$. The number of additions and multiplications required in the computation of PAPR of an OFDM symbol of length $N$ are $4N-1$ and $4N+1$, respectively. The null detection operation of the proposed scheme is slightly more complex than the other schemes. The required number of additions and multiplications in the detection process of the proposed scheme are $\frac{(4R-2P^{-1}+1)}{2P}$ and $N_d$, respectively.  A comparative complexity of the original scheme~\cite{wong1}, Ahmed et.~al.~\cite{sabbir2}, and proposed scheme is given in Table~\ref{tabl2}. For $P=2$, while the maximum number of additions and multiplications needed in the original method are 847314 and 595374, respectively, the corresponding numbers are 16640 and 11726 for our proposed scheme, with 98.04\% and 98.03\% reduction of the total additions and multiplications, respectively. It is worth noting that the complexity of our proposed scheme is 66.10\% lower than that of the improved scheme presented by Ahmed et.~al.~in~\cite{sabbir2}.
    
    In the system like~\cite{wong2}, the no-SI advantage is compromised to reduce the system complexity and improve BER. If we consider our proposed scheme in a 'non-blind' scenario, it remains more spectrally efficient compared to~\cite{wong2}. If $P$ data-null subcarriers switchings are done, the indexes of the $P$ SNSs are needed to send as SI in~\cite{wong2}. For a system of $N$ total subcarriers, the number of bits needed to send as SI becomes $Plog_2(N)$. In contrast, the non-blind scenario of the proposed system requires to send only $log_2(N)$ bits irrespective of the number of data-null switchings. The reason is that due to the equidistance property, only the index of the first SNS are required to send and all indexes of the subsequent SNSs can be derived from this received index at the receiver using the equidistance property.

\section{Conclusion}
 Due to the high complexity and poor BER of the conventional data-null subcarrier switching based PAPR reduction scheme, we propose an improved PAPR reduction scheme, with low computational overhead, which can decrease such effects dramatically. Compromising only 9.5\% of the PAPR reduction capability at $10^{-3}$ CCDF level, our scheme reduces the system complexity by more than 98\% and achieves up to 6 dB SNR gain.

\ifCLASSOPTIONcaptionsoff
  \newpage
\fi

\end{document}